\documentclass[conference]{IEEEtran}
\IEEEoverridecommandlockouts
\usepackage{amsmath,amssymb,amsfonts}
\usepackage{mathtools}
\usepackage{algorithmic}
\usepackage{graphicx}
\usepackage{textcomp}
\usepackage{xcolor}
\usepackage{hyperref}
\usepackage{svg}

\def\BibTeX{{\rm B\kern-.05em{\sc i\kern-.025em b}\kern-.08em
    T\kern-.1667em\lower.7ex\hbox{E}\kern-.125emX}}

\DeclareMathOperator*{\argmax}{arg\,max}

\usepackage[
  style=ieee,
  ]{biblatex}
\AtBeginBibliography{\footnotesize}

\begin{document}

\title{Packet-Level Complex CIR Tracking for Communication-Native mmWave Sensing: An 802.11ad Testbed}

\author{
\IEEEauthorblockN{
Raquel Marina Noguera Oishi$^{\ddagger *}$,
Haoqiu Xiong$^{\ddagger *\,\dagger}$,
Hany Assasa$^{*}$,
Sofie Pollin$^{*\,\dagger}$
\thanks{$^{\ddagger}$These authors contributed equally to this work.}%
}
\IEEEauthorblockA{
  $^{*}$Department of Electrical Engineering (ESAT), KU Leuven, Leuven, Belgium 
$^{\dagger}$Imec, Leuven, Belgium \\
\{raquelmarina.nogueraoishi, haoqiu.xiong, hany.assasa, sofie.pollin\}@kuleuven.be
}
}
\maketitle

\begin{abstract}
Commercial millimetre-wave (mmWave) Wi-Fi platforms based on IEEE 802.11ad/ay have demonstrated sensing capabilities, but they provide limited visibility into the packet-level measurements and synchronisation mechanisms that determine temporal coherence. We present a programmable 60~GHz IEEE 802.11ad testbed that exposes the complete complex 128-tap channel impulse response (CIR) for every detected packet. This access exposes packet-to-packet delay shifts and common phase drift that commercial-transceiver studies generally do not characterise. We develop a delay-profile alignment method and define a static-reference relative CIR phase that cancels common drift while preserving target-induced variation. We use 30~s respiration sensing to evaluate long-term phase stability for sub-wavelength motion tracking. Across 39 recordings, median absolute waveform correlation increases from 0.41 (timestamp regularised) to 0.52 (delay aligned) and 0.61 (relative-phase calibrated); breathing-rate mean absolute error decreases from 5.1 to 4.1 and 3.6~breaths/min. The results demonstrate controlled characterisation and stabilisation of communication-native WiGig CIR measurements on a fully programmable physical layer.
\end{abstract}

\begin{IEEEkeywords}
IEEE 802.11ad, WiGig sensing, channel impulse response, temporal coherence, mmWave testbed, vital signs
\end{IEEEkeywords}

\section{Introduction}
Wi-Fi sensing reuses the channel estimates that wireless local-area network receivers already compute for communication. Temporal changes in channel state information (CSI) or the channel impulse response (CIR) have enabled localisation, activity recognition, and contactless physiological monitoring \cite{wifi_csi_survey, sub6_respiratory_sensing}. These estimates are produced by a packet receiver rather than a dedicated sensing instrument. Their slow-time evolution therefore depends on packet detection, timing and frequency synchronisation, channel estimation, and the receiver interface through which the measurements are exposed.

WiGig extends this sensing approach to the 60~GHz band. IEEE 802.11ad/ay radios combine GHz bandwidth, millimetre-scale wavelength, and electronically steerable arrays, providing fine delay discrimination and phase sensitivity to small motion \cite{survey_mmWave_sensing, dmgsensing}. Existing work uses these properties for localisation and respiration sensing \cite{multiperson_sensing_11ay, WiGig_sensing}. ViMo \cite{ViMo}, for example, uses a commercial IEEE 802.11ad chipset to obtain CIR measurements at 20~Hz. It applies on-chip beam scanning, detects people in angle-range cells, and estimates their respiration and heart rates from the temporal evolution of CIR phase \cite{ViMo}. This result establishes that WiGig CIR phase contains useful physiological-motion information.

ViMo's processing begins with chipset-generated CIR measurements. Its signal model assumes that a reflection remains within one delay tap, treats the complex channel gain as time invariant for small subject motion, and absorbs a common phase shift into that gain \cite{ViMo}. The paper then focuses on beamformed target detection and vital-sign extraction; it does not report the complete delay profile and its stability, missing-snapshot behaviour, or the packet-level timing and hardware implementations behind the CIR sequence. We therefore address a complementary question: \emph{do communication-native CIR measurements remain aligned in delay and coherent in phase across packets?} These properties are essential for long-duration Doppler analysis or sub-wavelength motion sensing.

This paper examines the packet-level measurement process directly. Our programmable IEEE 802.11ad receiver exports all 128 complex CIR taps for every detected packet. The resulting sequence exposes missed detections, shifts of the complete delay profile, and phase drift common to all paths. We characterise these effects under controlled cabled and over-the-air conditions, then compensate them through delay-profile alignment and static-reference phase calibration. Respiration is used as a demanding validation case: a 30~s record must preserve the phase evolution of one target-associated tap while the chest moves by only a fraction of the 4.96~mm carrier wavelength.

Our contributions are:
\begin{itemize}
    \item A programmable IEEE 802.11ad physical layer (PHY), built on RFSoC platforms and 60~GHz front-ends, that extends the MIMORPH platform \cite{MIMORPH} to export the full complex 128-tap CIR, rather than only the strongest tap, for every detected packet. Packet training and beam control remain configurable.
    \item An experimental characterisation of packet-level CIR coherence, identifying quantised delay-profile shifts associated with packet detection and multipath, together with common phase drift from independent radio references.
    \item A stabilisation chain that aligns the delay profile and forms a static-reference relative CIR phase to cancel common phase drift while preserving target motion. Across 39 respiration recordings, median absolute waveform correlation increases from 0.41 (timestamp regularised) to 0.52 (delay aligned) and 0.61 (relative-phase calibrated), while breathing-rate mean absolute error decreases from 5.1 to 4.1 and 3.6~breaths/min.
\end{itemize}

The remainder of the paper is organised as follows: Section~\ref{sec:testbed_architecture} presents the testbed architecture, digital implementation, and characterisation. Section~\ref{sec:experimental_setup} describes the validation experiment. Section~\ref{sec:processing} formulates the CIR stabilisation and respiration-extraction chain, and Section~\ref{sec:results} presents the experimental results. Finally, Section~\ref{sec:conclusions} concludes the paper.

\section{Testbed architecture and characterisation}\label{sec:testbed_architecture}
\subsection{Hardware components}
\begin{figure}[t]
  \vspace{-10pt}
    \centering
    \includegraphics[width=\columnwidth]{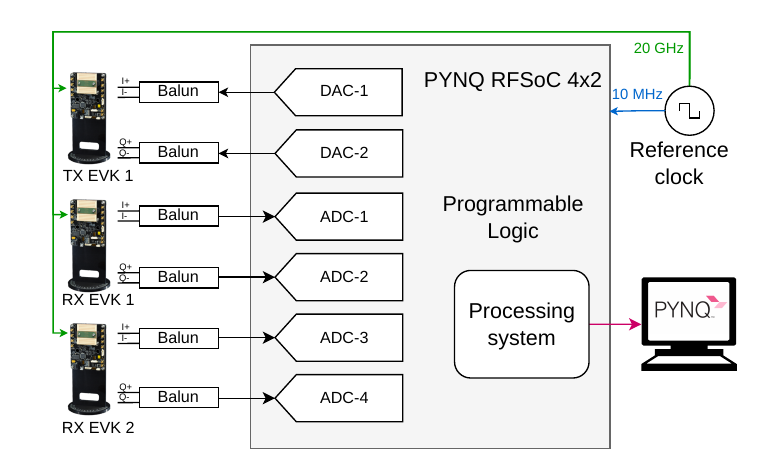}
    \caption{Architecture of one testbed radio node.}
    \label{fig:tb_CIR}
\end{figure}

The testbed comprises modular radio nodes, each formed by an RFSoC 4x2 board and Sivers evaluation-kit (EVK) front-ends that convert between baseband and the 60~GHz band (Fig.~\ref{fig:tb_CIR}). A node performs complex waveform generation, digital-to-analogue conversion (DAC), and analogue-to-digital conversion (ADC) on the RFSoC. Its two DAC channels and four ADC channels support one transmit front-end and up to two receive front-ends, because each Sivers front-end requires an in-phase/quadrature pair. The modular architecture supports monostatic, bistatic, and multistatic configurations.
The data converters operate at 3.52~GS/s, supporting a signal bandwidth of up to 1.76~GHz. An external generator supplies a 10~MHz reference to each RFSoC board and a 20~GHz reference from which each Sivers front-end synthesises the 60~GHz carrier. Separate generators can provide independent timing and frequency references to different nodes. The Sivers EVK steers its integrated patch-array beam in azimuth through a general-purpose input/output (GPIO) interface.

The RFSoC board connects to a host computer through Ethernet and runs a Jupyter server on the PYNQ framework. Python libraries configure the processing system and its FPGA intellectual-property blocks at run time.

\subsection{Digital baseband implementation}
\begin{figure}[htbp]
    \centering
    \includegraphics[width=0.45\textwidth]{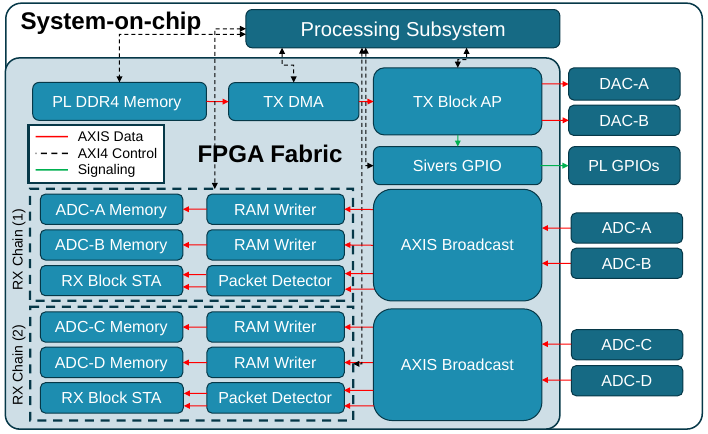}
    \caption{Design of the processing blocks in the programmable logic.}
    \label{fig:digital_design}
\end{figure}

The FPGA design in Fig.~\ref{fig:digital_design} contains one transmit path and two parallel receive paths, with run-time parameters controlled from software. It builds on open-source hardware accelerators from MIMORPH \cite{MIMORPH}, including the packet detector, transmit access point (TX AP), receive station (RX STA), and Sivers-controller blocks. We adapt these blocks to the sampling rate, radio configuration, and CIR-export requirements of this testbed.

\begin{figure*}[ht]
    \centering
    \includegraphics[width=0.8\textwidth]{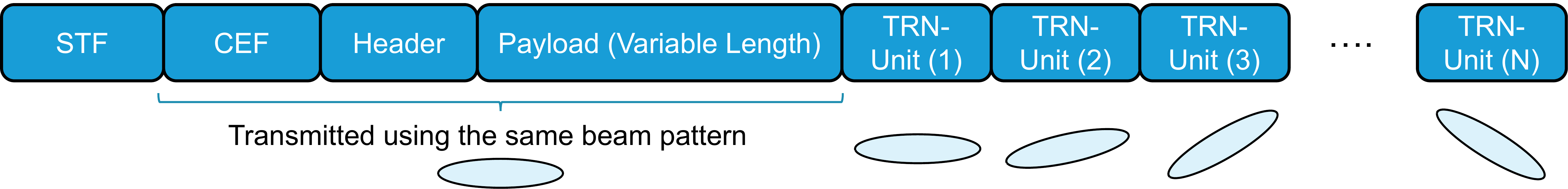}
    \caption{IEEE 802.11ad/ay packet structure with training fields.}
    \label{fig:11adpacket}
\end{figure*}

The TX block \cite{MIMORPH} generates an 802.11ad/ay-compliant single-carrier PHY protocol data unit entirely in hardware. The frame in Fig.~\ref{fig:11adpacket} contains a short training field for packet detection, automatic gain control, and frequency synchronisation; a channel-estimation field; a PHY header; a user-defined payload; one or more training (TRN) units for beam training or channel sounding; and a programmable idle interval. The Sivers GPIO block controls the transmit beam through the EVK control pins. At the start of the TRN field, the TX AP block generates a trigger that can advance the beam index for each TRN unit.

The receiver provides two parallel data paths. The first stores raw in-phase/quadrature samples for detailed offline analysis; because of the data volume, the available on-board memory holds approximately 64,000 complex samples. The second feeds ADC samples directly into the packet-detection and CIR-estimation pipeline:
\begin{itemize}
    \item Packet detector: this block continuously monitors the incoming stream and flags the arrival of a packet by correlating against the periodic Golay preamble, triggering the rest of the receive chain.
    \item RX STA block: this block estimates the channel over each TRN unit by correlating the received samples with the complementary Golay sequences defined by IEEE 802.11ad. The correlator outputs are coherently combined across repeated training sequences to improve the signal-to-noise ratio, producing one 128-tap CIR estimate per TRN unit. The original MIMORPH implementation \cite{MIMORPH} exports only the strongest tap. We modify it to export all 128 complex taps. With one TRN unit per packet, the configuration used in this paper yields one CIR snapshot per detection at a nominal rate of approximately 453~packets/s. The rate is configurable through the frame fields and inter-packet idle interval.
\end{itemize}


\subsection{Testbed characterisation: CIR peak-tap behaviour}
ViMo treats chipset-generated, beamformed CIR as input to adaptive object detection, motion classification, spatial clustering, and vital-sign estimation \cite{ViMo}; our objective is to characterise the CIR production process itself. Controlled cabled and over-the-air (OTA) tests therefore isolate how clock synchronisation and propagation affect the estimator independently of vital-sign extraction.

\begin{figure}[t]
    \centering
    \includegraphics[width=\columnwidth]{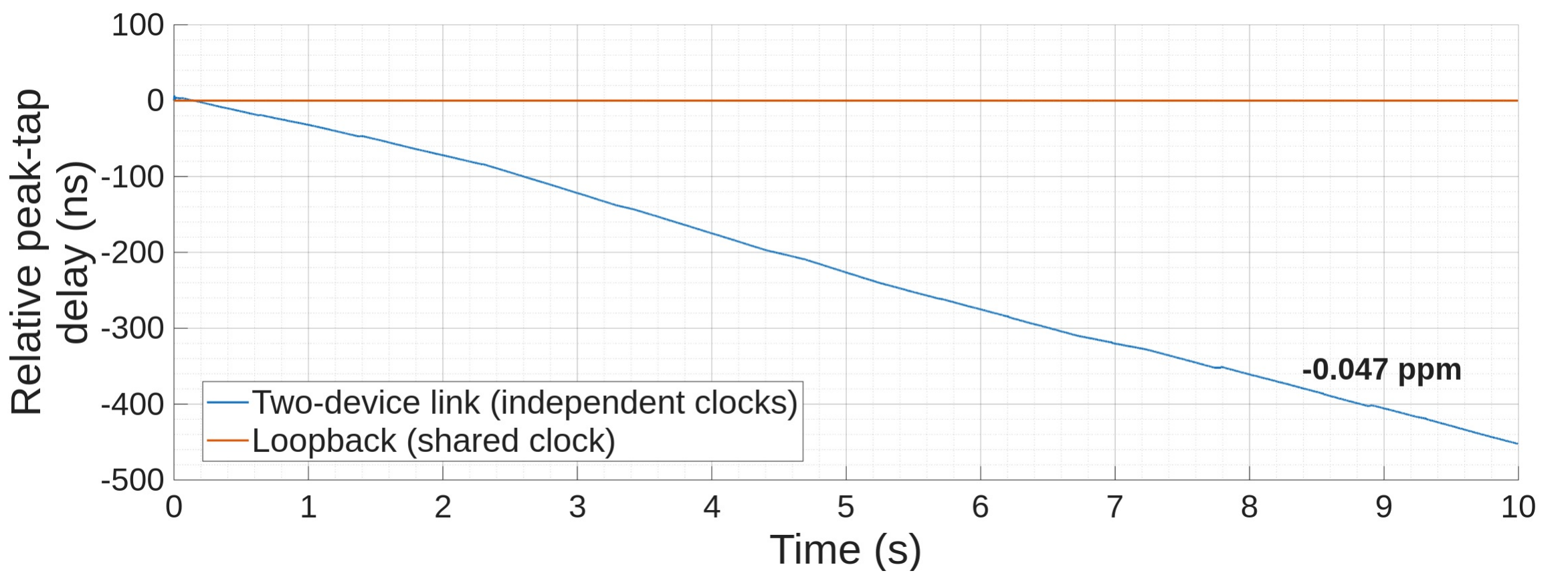}
   \caption{Effect of clock offset on the estimated CIR delay.}
    \label{fig:clock_offset}
\end{figure}

\textbf{Effect of clock synchronisation.} A cabled experiment isolates clock effects from the wireless channel. We compare a synchronised loopback, in which the transmitter and receiver share a reference clock, with a two-device link driven by independent clock generators. Fig.~\ref{fig:clock_offset} shows the dominant-tap delay over a 10~s capture. With independent clocks, a sampling-rate offset produces a delay drift whose fitted slope corresponds to approximately 0.047~ppm. The shared-clock loopback keeps the dominant tap at a constant delay. Distributed and bistatic sensing with separate nodes therefore requires explicit sampling-clock synchronisation or compensation.

\begin{figure}[t]
    \centering
    \includegraphics[width=0.45\textwidth]{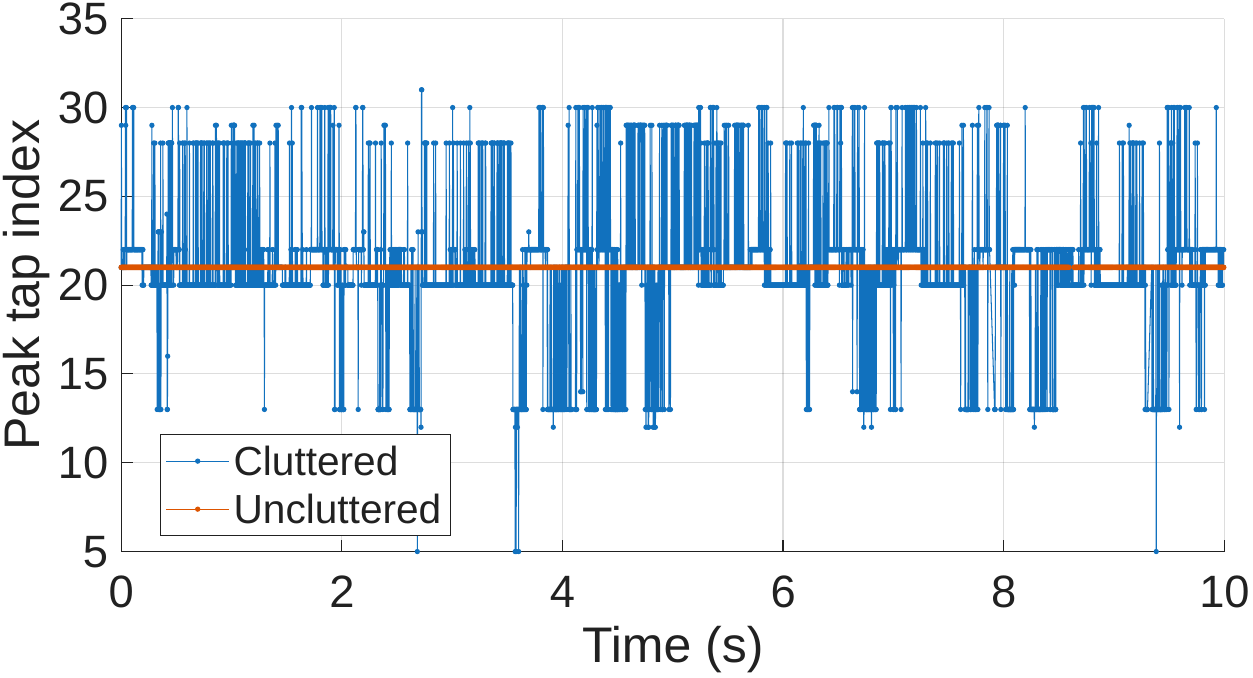}
   \caption{Effect of the wireless channel on the CIR peak tap. The cluttered link shows variations in the peak tap due to multipath, while the uncluttered link maintains a stable peak tap.}
    \label{fig:multipath_effect}
\end{figure}

\textbf{Effect of the wireless channel.} The second experiment evaluates the CIR estimator under two propagation conditions using an OTA loopback configuration. Fig.~\ref{fig:multipath_effect} compares an uncluttered link with a cluttered link containing additional reflectors. The dominant tap remains stable in the uncluttered link but changes in the cluttered link. Under multipath, ambiguities in packet detection and CIR estimation can cause the receiver to select different timing references, shifting the complete delay profile. The subsequent sensing experiments therefore align each CIR profile before extracting motion-induced phase.

\section{Experimental setup}\label{sec:experimental_setup}
\begin{figure}[htbp]
\centerline{\includegraphics[width=0.4\textwidth]{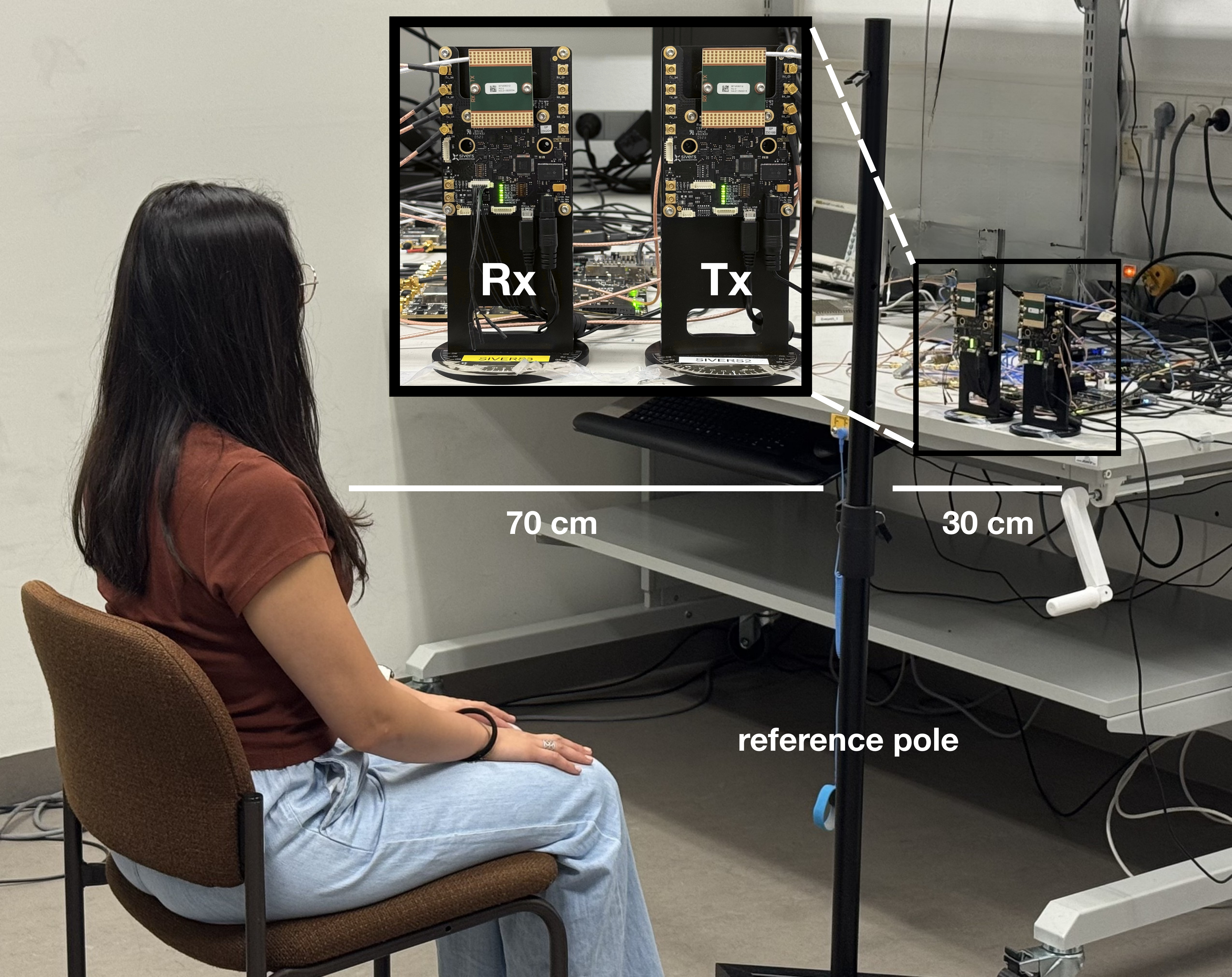}}
\caption{Experimental setup for vital-sign sensing.}
\label{fig:setup}
\end{figure}

The validation experiment uses a quasi-monostatic architecture: one RFSoC 4x2 board drives two co-located Sivers front-ends, one transmitting and one receiving. A static pole in the scene provides a phase-reference CIR tap (Fig.~\ref{fig:setup}).

In each experiment, a participant is seated on a chair behind the reflector, so the receiver observes two distinct propagation paths: a static path via the reflector and a dynamic path via the participant. This geometry allows the phase difference between the static-reference tap and the target-associated tap to be extracted directly from the CIR \cite{JUMP}. This validation does not exploit the beamforming capability of the front-ends; it uses a fixed broadside beam and one TRN unit per IEEE 802.11ad packet.

Each recording lasts 30~s. The transmitter sends packets continuously, while the receiver detects packets, estimates one CIR per detected packet, and stores the CIR sequence for offline analysis.

The subject wears a Zephyr BioHarness chest belt that records respiration at 25~Hz. The campaign contains 39 recordings from six subjects (four males and two females). Each subject is seated 1~m from the node, with the reflector at 30~cm, and adopts one of three body orientations ($0^{\circ}$ or $\pm45^{\circ}$). The recordings include normal breathing, controlled changes in breathing rate, and deliberate apnoea.

The Social and Societal Ethics Committee at KU Leuven approved the protocol for the human-subject experiments.

\section{CIR stabilisation and respiration extraction}\label{sec:processing}
Respiration changes the propagation delay, and hence the phase, of the CIR tap associated with the torso. The processing chain converts this phase variation into a displacement waveform. Two testbed-specific effects must first be removed: every CIR snapshot contains a residual common phase from the radio clocks, and packet detection under multipath introduces a timing offset that can shift the complete delay profile.

\subsection{Signal model}
Let $h_n[k]$ denote the CIR estimate from the single TRN unit of the $n$-th detected packet, hereafter a \emph{CIR snapshot}. The slow-time packet index is $n=0,\dots,N-1$, the nominal packet rate is $f_{\mathrm p}$, and the delay-tap index is $k=0,\dots,K-1$, with $K=128$. Taps are separated by $\Delta\tau=1/B$ for $B=1.76$~GHz. Under the quasi-monostatic two-way propagation assumption, this delay spacing corresponds to the radial-range resolution $\Delta r=c/(2B)=8.5$~cm. Modelling the scene as $P$ specular paths,
\begin{equation}
\begin{split}
h_n[k]=e^{j\varphi_n}\sum_{p=1}^{P}\alpha_{p,n}\,e^{-j2\pi f_c\tau_{p,n}}\,
g\!\left(k\Delta\tau-\tau_{p,n}-\delta_n\right)\\
+\;w_n[k],
\end{split}
\label{eq:model}
\end{equation}
where $\alpha_{p,n}$ is the complex scattering coefficient of path $p$, excluding the propagation phase $e^{-j2\pi f_c\tau_{p,n}}$ written explicitly in \eqref{eq:model}; $\tau_{p,n}$ is the propagation delay; $g(\cdot)$ is the Golay autocorrelation pulse; $f_c=60.48$~GHz is the carrier frequency (wavelength $\lambda=4.96$~mm); and $w_n[k]$ is noise. Two platform artefacts appear in \eqref{eq:model}:
\begin{itemize}
    \item $\varphi_n$, the phase accumulated between packets because of residual clock offsets between the transmit and receive chains. It is \emph{common to all paths} in $h_n[\cdot]$.
    \item $\delta_n$, the residual timing offset left by the packet detector. It shifts the entire delay profile.
\end{itemize}

The quantity of interest is carried by the target path. With the torso at nominal range $R_{\mathrm t}$ and chest displacement $d(t)$, the delay is $\tau_{\mathrm t,n}=2\left(R_{\mathrm t}+d(t_n)\right)/c$ and the phase of the corresponding tap is
\begin{equation}
\phi_n=-\frac{4\pi}{\lambda}\left(R_{\mathrm t}+d(t_n)\right)+\varphi_n+\mathrm{const}.
\label{eq:phase}
\end{equation}
The motion-dependent phase in \eqref{eq:phase} can be tracked reliably only after correcting the common phase $\varphi_n$ and the delay-profile shift caused by $\delta_n$.

\subsection{Slow-time regularisation and CIR alignment}
Missed detections make the timestamps $t_n$ nonuniform. Taking each recording's median packet interval as its nominal interval, we estimate that 2.8\% of the expected snapshots are missing across the 39 recordings. ViMo reports a 20~Hz sampling rate and formulates adjacent CIR samples at a fixed interval, but it does not discuss missing snapshots or irregular-timestamp handling \cite{ViMo}. We linearly interpolate the real and imaginary CIR components from their recorded timestamps onto a uniform grid before applying filters specified in hertz.

The timing offset $\delta_n$ is estimated as an integer tap shift. A per-packet gain change would bias a direct amplitude match, so alignment uses the log-magnitude profile. With $\epsilon$ a small floor, define the packet profile and its median template as
\begin{equation}
\ell_n[k]=\log\left(\left|h_n[k]\right|+\epsilon\right),\;
\bar{\ell}[k]=\operatorname*{median}_{n}\ell_n[k].
\end{equation}
The shift and the aligned CIR are
\begin{equation}
\hat{q}_n=\argmax_{q\in\{-L,\ldots,L\}}\,
\rho_k\!\left(\ell_n[k+q],\bar{\ell}[k]\right),
\label{eq:align}
\end{equation}
\begin{equation}
h^{\mathrm a}_n[k]=h_n[k+\hat{q}_n],
\end{equation}
where tap indices are circular, $\rho_k(\cdot,\cdot)$ is the Pearson correlation across $k$, and $L=20$ taps. Ties favour the smallest $|q|$. The median template is robust when slips affect fewer than half of the packets. Equation \eqref{eq:align} is applied twice, recomputing the template after the first pass. The reference and target taps, $\hat{k}_{\mathrm r}$ and $\hat{k}_{\mathrm t}$, are then selected as the two dominant peaks of the aligned mean profile.

\begin{figure}[t]
\centerline{\includegraphics[width=\columnwidth]{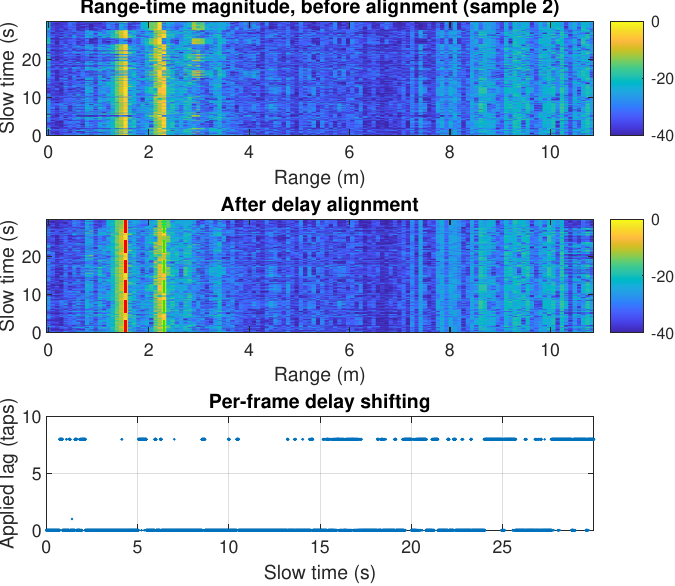}}
\caption{Range--time magnitude of one 30~s recording before (top) and after (middle) delay alignment, and the per-packet shift applied (bottom). The dashed lines mark the reference and target taps. The slip is quantised to 8 taps and affects 26.9\% of the packets in this recording.}
\label{fig:alignment}
\end{figure}

\subsection{Relative CIR phase and displacement extraction}
\vspace{-1pt}
We define the \emph{relative CIR phase} by referencing every aligned tap to the static-reflector tap within the same snapshot,
\begin{equation}
\begin{aligned}
\tilde{h}_n[k]&=h^{\mathrm a}_n[k]\left(h^{\mathrm a}_n\!\left[\hat{k}_{\mathrm r}\right]\right)^{\!*},\\
\theta_n^{\mathrm{rel}}[k]&\triangleq\angle\tilde{h}_n[k].
\end{aligned}
\label{eq:refcal}
\end{equation}
Both taps are acquired in the same snapshot and contain the same packet-dependent phase $\varphi_n$; the conjugate product subtracts the reference phase. Hence, any phase variation common to both taps is cancelled, while variation caused by target motion relative to the static reflector is preserved. Because the reflector is stationary, its propagation phase is constant apart from noise, so $\theta_n^{\mathrm{rel}}[\hat{k}_{\mathrm t}]$ contains the motion-dependent term in \eqref{eq:phase}, up to a constant offset and residual noise.

At 60.48~GHz, a 1~cm displacement induces approximately an $8\pi$ phase rotation, so the target relative-phase sequence $\theta_n^{\mathrm{rel}}[\hat{k}_{\mathrm t}]$ is unwrapped from successive phase increments. An increment is retained only when both adjacent magnitudes exceed 5\% of the sequence's 95th-percentile magnitude and its absolute phase change does not exceed $\pi/2$; rejected increments contribute zero. This gate prevents a low-amplitude sample or an anomalously large increment from creating a persistent $2\pi$ error. Let $\psi_n$ denote the resulting unwrapped phase. After linear detrending and zero-phase third-order Butterworth filtering over $[0.1,0.5]$~Hz, inversion of \eqref{eq:phase} gives
\begin{equation}
\hat{d}_n=-\frac{\lambda}{4\pi}\,\psi^{\mathrm{bp}}_n.
\label{eq:disp}
\end{equation}

Breathing rate is then estimated from the extracted displacement waveform by counting observed cycles.

\section{Temporal-coherence validation}\label{sec:results}
We apply the chain of Section~\ref{sec:processing} to all 39 recordings. One broadside ($0^{\circ}$) normal-breathing recording is examined in detail before the aggregate results are presented.

\begin{figure}[t]
\centerline{\includegraphics[width=0.9\columnwidth]{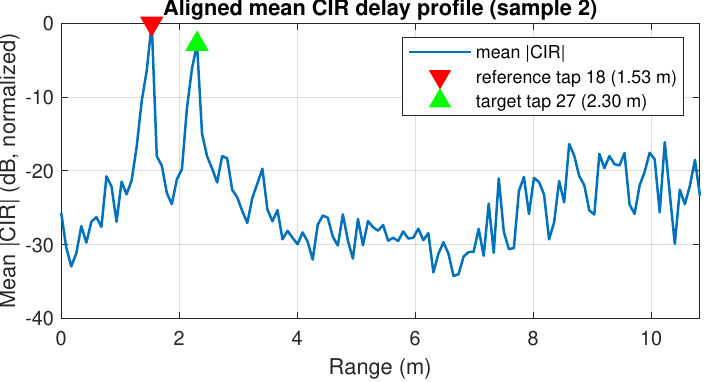}}
\caption{Mean aligned CIR delay profile of one recording, with the automatically selected reference (static reflector) and target (torso) taps.}
\label{fig:cir_profile}
\end{figure}

Fig.~\ref{fig:cir_profile} shows the mean aligned delay profile of the example recording. Two peaks dominate: the static reflector at tap 18 and the seated subject at tap 27, corresponding to a range separation of 0.77~m from the static reflector. This agrees with the 0.70~m spacing between the pole and the chair to within one delay tap (8.5~cm). The absolute delays include a fixed system offset through the cabling and the processing chain, so only the difference between taps is geometrically meaningful. Over 30~s, the receiver detects 13\,368 CIR snapshots; timestamp regularisation reconstructs a 13\,539-point grid at $f_{\mathrm p}=451.32$~Hz, corresponding to 1.3\% missing snapshots. Of the detected profiles, 26.9\% are affected by the delay slip corrected in Fig.~\ref{fig:alignment}.

\begin{figure}[!t]
\centerline{\includegraphics[width=\columnwidth]{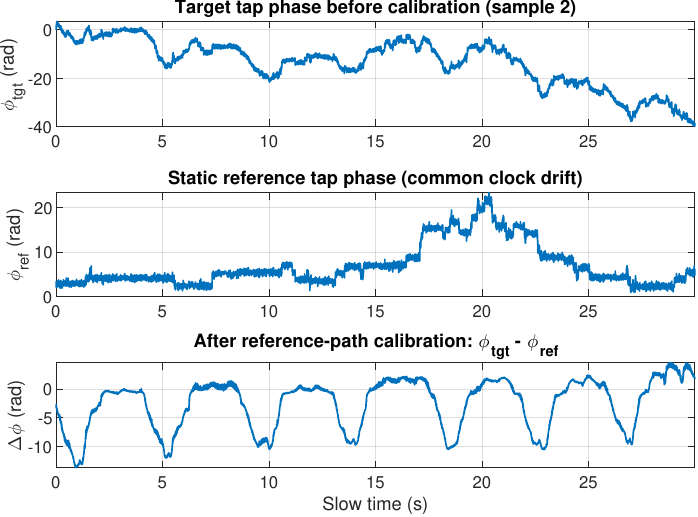}}
\caption{Relative CIR phase from static-reference calibration \eqref{eq:refcal}. The target phase (top) and static-reference phase (middle) share residual clock drift; their relative phase (bottom) reveals the respiration cycles.}
\label{fig:refcal}
\end{figure}

Fig.~\ref{fig:refcal} illustrates the effect of static-reference phase calibration. The target and static-reference phases share excursions of tens of radians over 30~s. Their relative phase suppresses this common component before unwrapping and filtering while retaining the target-path modulation produced by respiration. This is the practical justification for placing a static reflector in the scene.

\begin{figure}[!t]
\centerline{\includegraphics[width=0.9\columnwidth]{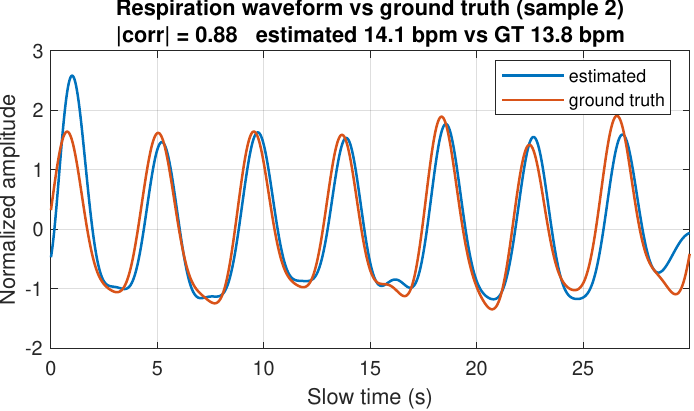}}
\caption{Respiration waveform estimated from a single CIR tap, compared with the chest-belt reference over the same 30~s window. Both waveforms are normalised; the estimate reproduces the individual breathing cycles, including the asymmetry between inhalation and exhalation.}
\label{fig:resp_gt}
\end{figure}

The displacement waveform from \eqref{eq:disp} is compared with the chest-belt reference in Fig.~\ref{fig:resp_gt}. The estimate follows the individual cycles, including the asymmetry between inhalation and exhalation, and achieves an absolute waveform correlation of 0.88 over the 30~s window. Its estimated breathing rate is 14.1~breaths/min, compared with 13.8~breaths/min for the reference, an error of 0.3~breaths/min. The recovered peak-to-peak chest excursion is 5.9~mm.

Across recordings, waveform agreement is quantified by $|\rho_{\mathrm{resp}}|$, the absolute Pearson correlation between the estimated and chest-belt respiration waveforms after detrending and filtering over $[0.1,0.5]$~Hz. The absolute value is used because radar phase and chest-belt displacement have an arbitrary relative sign. For recording $i$, the absolute rate error is $e_i=|\hat r_i-r_i^{\mathrm{ref}}|$. Table~\ref{tab:ablation} reports the median $|\rho_{\mathrm{resp}}|$ and the mean absolute rate error (MAE), $M^{-1}\sum_{i=1}^{M}e_i$, over the $M=39$ recordings.

\begin{table}[t]
\caption{Stagewise CIR stabilisation ablation over all 39 recordings. The rate MAE is reported in breaths/min.}
\label{tab:ablation}
\centering
\footnotesize
\setlength{\tabcolsep}{4pt}
\begin{tabular}{lcc}
\hline
CIR stage & Median $|\rho_{\mathrm{resp}}|$ & Rate MAE \\
\hline
Timestamp regularised & 0.41 & 5.1 \\
Delay aligned & 0.52 & 4.1 \\
Relative-phase calibrated & \textbf{0.61} & \textbf{3.6} \\
\hline
\end{tabular}
\end{table}

Table~\ref{tab:ablation} separates the two platform-specific corrections. All stages use the same uniform slow-time grid, the same target tap selected from the aligned mean profile, identical reliability thresholds, and the same respiration filter and evaluation procedure. The timestamp-regularised stage applies neither CIR correction. The delay-aligned stage adds \eqref{eq:align}, and the relative-phase-calibrated stage additionally applies \eqref{eq:refcal}. Delay alignment increases the median absolute waveform correlation from 0.41 to 0.52 and reduces the breathing-rate MAE from 5.1 to 4.1~breaths/min. Relative-phase calibration further raises the correlation to 0.61 and reduces the rate MAE to 3.6~breaths/min.

For the 13 normal-breathing broadside recordings, the three median correlations are 0.58, 0.70, and 0.77, while the corresponding mean absolute rate errors are 3.87, 2.86, and 1.47~breaths/min. Performance is lower for the 12 normal-breathing recordings at $\pm45^\circ$: after relative-phase calibration, the median correlation is 0.31 and the mean absolute rate error is 6.28~breaths/min.

\section{Conclusion}\label{sec:conclusions}
We presented a programmable 60~GHz IEEE 802.11ad testbed that exports the complete complex 128-tap CIR of every detected packet. Direct access reveals delay-profile shifts associated with packet detection and multipath, together with phase drift common to the radio chains. Delay alignment and the proposed static-reference relative CIR phase increase median respiration-waveform correlation from 0.41 after timestamp regularisation to 0.52 and 0.61, while reducing breathing-rate mean absolute error from 5.1 to 4.1 and 3.6~breaths/min. Respiration validates coherent tracking over 30~s at sub-wavelength scale; the central contribution is the characterisation and relative-phase stabilisation of communication-native CIR measurements.
 
Future work targets inter-node carrier-frequency and sampling-rate offset correction, beam-steered and multi-node sensing, more robust static-reference selection, and fusion across the multiple range bins occupied by the torso. These capabilities will extend the platform from the present fixed-beam validation to multi-target and long-duration sensing experiments.

\section*{Acknowledgements}
This work was supported by the Horizon Europe Research and Innovation Programme under Grant 101192521 (MultiX).

\section*{References}
\printbibliography[notkeyword={ownwork}, heading=none]

@inproceedings{MIMORPH,
author = {Lacruz, Jesus O. and Ortiz, Rafael Ruiz and Widmer, Joerg},
title = {A real-time experimentation platform for sub-6 GHz and millimeter-wave MIMO systems},
year = {2021},
booktitle = {Proceedings of the 19th Annual International Conference on Mobile Systems, Applications, and Services},
pages = {427–439}}

@ARTICLE{dmgsensing,
  author={Blandino, Steve and Ropitault, Tanguy and da Silva, Claudio R. C. M. and Sahoo, Anirudha and Golmie, Nada},
  journal={IEEE Open Journal of Vehicular Technology}, 
  title={IEEE 802.11bf DMG Sensing: Enabling High-Resolution mmWave Wi-Fi Sensing}, 
  year={2023},
  volume={4},
  number={},
  pages={342-355},
  doi={10.1109/OJVT.2023.3237158}}

@ARTICLE{JUMP,
  author={Pegoraro, Jacopo and Lacruz, Jesus O. and Azzino, Tommy and Mezzavilla, Marco and Rossi, Michele and Widmer, Joerg and Rangan, Sundeep},
  journal={IEEE Transactions on Wireless Communications}, 
  title={JUMP: Joint Communication and Sensing With Unsynchronized Transceivers Made Practical}, 
  year={2024},
  volume={23},
  number={8},
  pages={9759-9775},
  doi={10.1109/TWC.2024.3365853}}

@ARTICLE{wifi_csi_survey,
  author={Ma, Yongsen and Zhou, Gang and Wang, Shuangquan},
  journal={ACM Computing Surveys},
  title={{WiFi} Sensing with Channel State Information: A Survey},
  year={2019},
  volume={52},
  number={3},
  pages={1-36},
  doi={10.1145/3310194}}

@ARTICLE{survey_mmWave_sensing,
  author={Zhang, Jia and Xi, Rui and He, Yuan and Sun, Yimiao and Guo, Xiuzhen and Wang, Weiguo and Na, Xin and Liu, Yunhao and Shi, Zhenguo and Gu, Tao},
  journal={IEEE Communications Surveys \& Tutorials}, 
  title={A Survey of mmWave-Based Human Sensing: Technology, Platforms and Applications}, 
  year={2023},
  volume={25},
  number={4},
  pages={2052-2087}}

@inproceedings{multiperson_sensing_11ay,
author = {Xiong, Haoqiu and Cui, Zhuangzhuang and Liu, Mingqing and Miao, Yang and Pollin, Sofie},
title = {Multi-person Localization and Respiration Sensing under IEEE 802.11ay Standard},
year = {2023},
booktitle = {Proceedings of the 3rd ACM MobiCom Workshop on Integrated Sensing and Communications Systems},
pages = {31–36}
}

@INPROCEEDINGS{WiGig_sensing,
  author={Xiong, Haoqiu and Cui, Zhuangzhuang and Miao, Yang and Pollin, Sofie},
  booktitle={ICASSP 2024 - 2024 IEEE International Conference on Acoustics, Speech and Signal Processing (ICASSP)}, 
  title={WiGig-based Joint Multi-Person Positioning and Respiration Sensing}, 
  year={2024},
  volume={},
  number={},
  pages={13321-13325}}

@ARTICLE{ViMo,
  author={Wang, Fengyu and Zhang, Feng and Wu, Chenshu and Wang, Beibei and Liu, K. J. Ray},
  journal={IEEE Internet of Things Journal},
  title={ViMo: Multiperson Vital Sign Monitoring Using Commodity Millimeter-Wave Radio},
  year={2021},
  volume={8},
  number={3},
  pages={1294-1307},
  doi={10.1109/JIOT.2020.3004046}}

@ARTICLE{sub6_respiratory_sensing,
  author={Xiong, Haoqiu and Beerten, Robbert and Zhang, Qing and Miao, Yang and Cui, Zhuangzhuang and Pollin, Sofie},
  journal={IEEE Journal on Selected Areas in Communications}, 
  title={Fundamentals and Experiments of Robust Respiration Sensing via Cell-Free Massive MIMO}, 
  year={2026},
  volume={44},
  number={},
  pages={959-974}}

\end{document}